\documentclass[12pt,journal,onecolumn]{IEEEtran}
\usepackage{cite}
\usepackage{amsmath,amssymb,amsfonts,amsthm}
\usepackage{algorithmic}
\usepackage{graphicx}
\usepackage{textcomp}
\usepackage{xcolor}
\usepackage{comment}
\usepackage{float}

\usepackage{setspace}
\usepackage{soul}
\usepackage{color}

\usepackage[caption=false,font=footnotesize]{subfig}
\DeclareMathOperator*{\esssup}{esssup}
\newtheorem{theorem}{Theorem}
\newtheorem{lemma}[theorem]{Lemma}
\newtheorem{corollary}[theorem]{Corollary}
\newtheorem*{theorem*}{Theorem}
\newtheorem*{lemma*}{Lemma}
\newtheorem*{corollary*}{Corollary}
\DeclareMathOperator*{\argmin}{argmin}
\usepackage{physics}
\usepackage{MnSymbol}
\usepackage[bb=boondox]{mathalfa}

\def\BibTeX{{\rm B\kern-.05em{\sc i\kern-.025em b}\kern-.08em
    T\kern-.1667em\lower.7ex\hbox{E}\kern-.125emX}}
\begin{document}

\title{Separating an Outlier from a Change}

\author{Deniz Sargun, C. Emre Koksal, \IEEEmembership{Senior Member, IEEE}%
\thanks{The  authors  are  with  the  Department  of  Electrical  and  Computer Engineering,  The  Ohio  State  University,  Columbus,  OH  43210  USA  (e-mail:\{sargun.1, koksal.2\}@osu.edu).}}


\maketitle

\setstcolor{red}

\begin{abstract}
We study the change detection problem with an unknown post-change distribution. Under this constraint, the unknown change in the distribution of observations may occur in many ways without much structure on the observations, whereas, before the change point, a false alarm (outlier) is highly structured, following a particular sample path. We first characterize these likely events for the deviation and propose a method to test the empirical distribution, relative to the most likely way for it to occur as an outlier. We benchmark our method with finite moving average (FMA) and generalized likelihood ratio tests (GLRT) under 4 different performance criteria including the run time time complexity. Finally, we apply our method on economic market indicators and climate data. Our method successfully captures the regime shifts during times of historical significance for the markets and identifies the current climate change phenomenon to be a highly likely regime shift rather than a random event.
\end{abstract}


\section{Introduction} \label{section: introduction}
Not every long-term deviation from the norm should be considered as a result of a change. Such occurrences may also be caused by rare events driven by the system and this distinction is important in many applications from model selection \cite{app1,app1-2} to hardware faults \cite{app2,app2-2}, security \cite{app3,app3-2} and even health \cite{app4,app4-2}. To that end, this paper focuses on problem instances of change detection \cite{poorBook} (also see \cite{Basseville} for a wide treatment of change detection and refer to chapter 8 for nonadditive change models and existing approaches) in a time-series with minimax cost with respect to change point and post-change distributions.

The quickest change detection problem with unknown post-change models has been studied widely, both in theory \cite{bin,asymptotically,minimax,unknown,nearly} and in application \cite{multidecision,optimalSeq} (see \cite{veer3} for more variants and applications). For some non-Bayesian quickest change detection problems where pre- and post-change distributions have finitely many alternatives it is known that a version of the cumulative sum algorithm (CUSUM) is optimal or asymptotically optimal \cite{alternatives,data,shewhart}. For other problems where the pre- or post-change distribution has infinitely many alternatives or is not parametrized, algorithms that form a likelihood ratio (ex. CUSUM) are not applicable. In this case, since the change in the distribution is arbitrary it is hard to perform optimally against all alternatives. A widely used method in the literature is to form a generalized likelihood ratio statistics over the set of alternative distributions and then use it as a drift term \cite{unknown,nuisance,data}, which can be computationally demanding or without any optimality guarantees. To address this, we propose a recursive \footnote{if the quasiconcave function \(q\) can be estimated recursively for the empirical probability mass function \(\hat{f}\)} and linear complexity online method, called the information projection test (IPT), that uses an initial convex decision region and modifies it to suppress most likely false alarms. IPT utilizes the separation between pre- and post-change distributions at the first level and compares the relative entropy of the empirical distribution with respect to the distribution of the most likely outlier at the second level. For the quickest change detection problem, we find bounds on the average run length (ARL) and the worst average detection delay (WADD) and prove its asymptotic optimality up to a multiplicative constant and share simulation results which empirically show that IPT performs similar to GLRT.

To motivate our approach, we first define IPT for fixed window size tests and bound the receiver operating characteristic (ROC) of IPT for the composite hypothesis testing problem. Then, we transition to the transient change detection problem to introduce the concept of delay and bound IPT's performance of detecting changes with bounded delay before studying Lorden's criteria. Again, we provide empirical comparisons with of FMA \cite{performance} and GLRT.

Another important metric considered is the computational complexity of the algorithms. There is already research on low complexity change point estimation techniques, two examples being changes in video content popularity \cite{youtube} and networks with propagating changes \cite{propagate}. Even with a similar detection performance, IPT has a few orders of lower complexity (\(99.86\%\) reduction in test time for an alphabet with \(\sim 6500\) letters and same order of samples) compared to GLRT. Thus, we claim our linear complexity algorithm achieves a better trade-off between computation time and performance compared to other methods in the literature.

Contributions of this paper can be summarized as follows:
\begin{itemize}
    \item We introduce a simple and novel algorithm for detecting changes with known pre-change distribution but unknown post-change distribution.
    \item We prove performance bounds in the non-asymptotic regime and prove asymptotic optimality.
    \item We empirically benchmark our method under (1) composite hypothesis testing problem, (2) worst case post-change distribution transient change detection problem and (3) worst case post-change distribution quickest change detection problem with Lorden's criteria.
    \item We study special cases to provide insight and application areas for our method.
    \item The time complexity of our approach is compared against existing methods theoretically and empirically.
    \item We draw new insights by applying our idea in analyzing economic and climate time series data to understand whether periods of long deviations from the norm are outliers or change. We identify shifts from the average behavior of market indices and company returns to show that our scheme successfully manages to detect the periods with regime changes like crises. One of the highlights of our analysis is that the climate change phenomenon that is observed in the last 30 years is highly unlikely to be an outlier, giving credence to the hypothesis that it is caused by exogenous factors.
\end{itemize}
Earlier version of this work has appeared in \cite{sargun}. In this version we include performance bounds for the transient and quickest change detection problems, include empirical results for the latter, increase the number of applications on real datasets and provide insightful special cases.

\section{Model and Problem Statement} \label{section: model}
Let \(\mathcal{A}=\left\{a_1,\dots,a_m\right\}\) be a finite alphabet and \(\mathcal{P}\) denote the probability simplex of probability mass functions (p.m.f.s) \(f\) over \(\mathcal{A}\). Consider \(X_1^\infty=\left(X_1,\dots,X_k,\dots\right)\), a sequence of observations, where, given change point \(t_1\), each random variable \(X_1^{t_1-1}\) is independent and identically distributed (i.i.d.) with known pre-change p.m.f. \(f_0\) and \(X_{t_1}^\infty\) are i.i.d. with the unknown post-change p.m.f. \(f_1\in\mathcal{P}_1\), independent of previous observations \(X_1^{t_1-1}\). Assume there exists \(q:\left(\mathcal{P},l_1\right)\to\mathbb{R}\) quasiconcave and Lipschitz continuous with Lipschitz constant \(L\), that satisfies \(q(f_0)=q_0<0<\underline{q}\leq q_1=q(f_1)\). In other words, \(\mathcal{P}_1\) is a subset of a closed and convex set \(q^{-1}\left(\left[\underline{q},\infty\right)\right)\) and \(f_0\notin q^{-1}\left(\left[\underline{q},\infty\right)\right)\). We use \(\hat{f}_{X_i^j}\) for the empirical p.m.f. of \(X_i^j\), ie. \(\hat {f}_{X_i^j}(x)=\frac{1}{j-i+1}\sum_{k=i}^j\mathbb{1}_{x_i}(x)\), and \(\mathcal{P}^n\) for the discrete set of empirical p.m.f.s that can be realized with \(n\) samples from \(\mathcal{A}\). Finally, \(I\left(f\middle\|f'\right)=\sum_a \log\frac{f(a)}{f'(a)}\) denotes the Kullback-Leibler (KL) divergence between distributions \(f\) and \(f'\) in nats and \(P_{f_1,t_1}\) denotes the probability law when post-change distribution is \(f_1\) and change time is \(t_1\). When change does not occur we denote \(P_{f_1,\infty}\) briefly as \(P_\infty\).

\begin{figure*}[!t]
\centering
\subfloat[Initial decision regions \((\Gamma_0,\Gamma_1)\), the distributions \(f_0\) and \(f_1\) and convex \(\left\{q(f)\geq\underline{q}\right\}\)]{\includegraphics[trim=160 0 230 0, clip,width=0.3\textwidth]{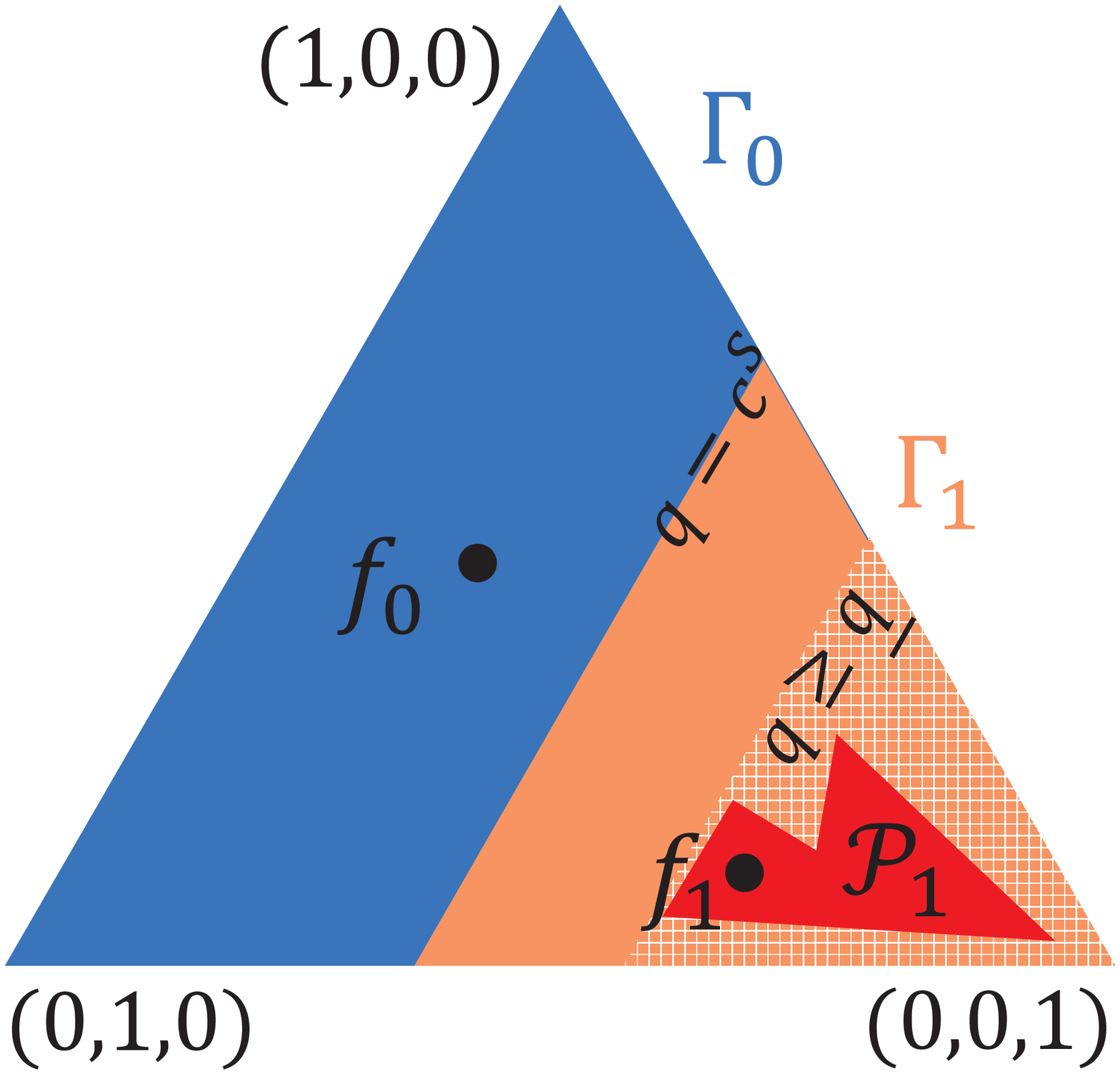}
\label{figure: simplex}}
\hfil
\subfloat[The I-projection \(f^*\in\left\{q(f)\geq c^S\right\}\) minimizes the KL divergence from \(f_0\)]{\includegraphics[trim=160 0 230 0, clip,width=0.3\textwidth]{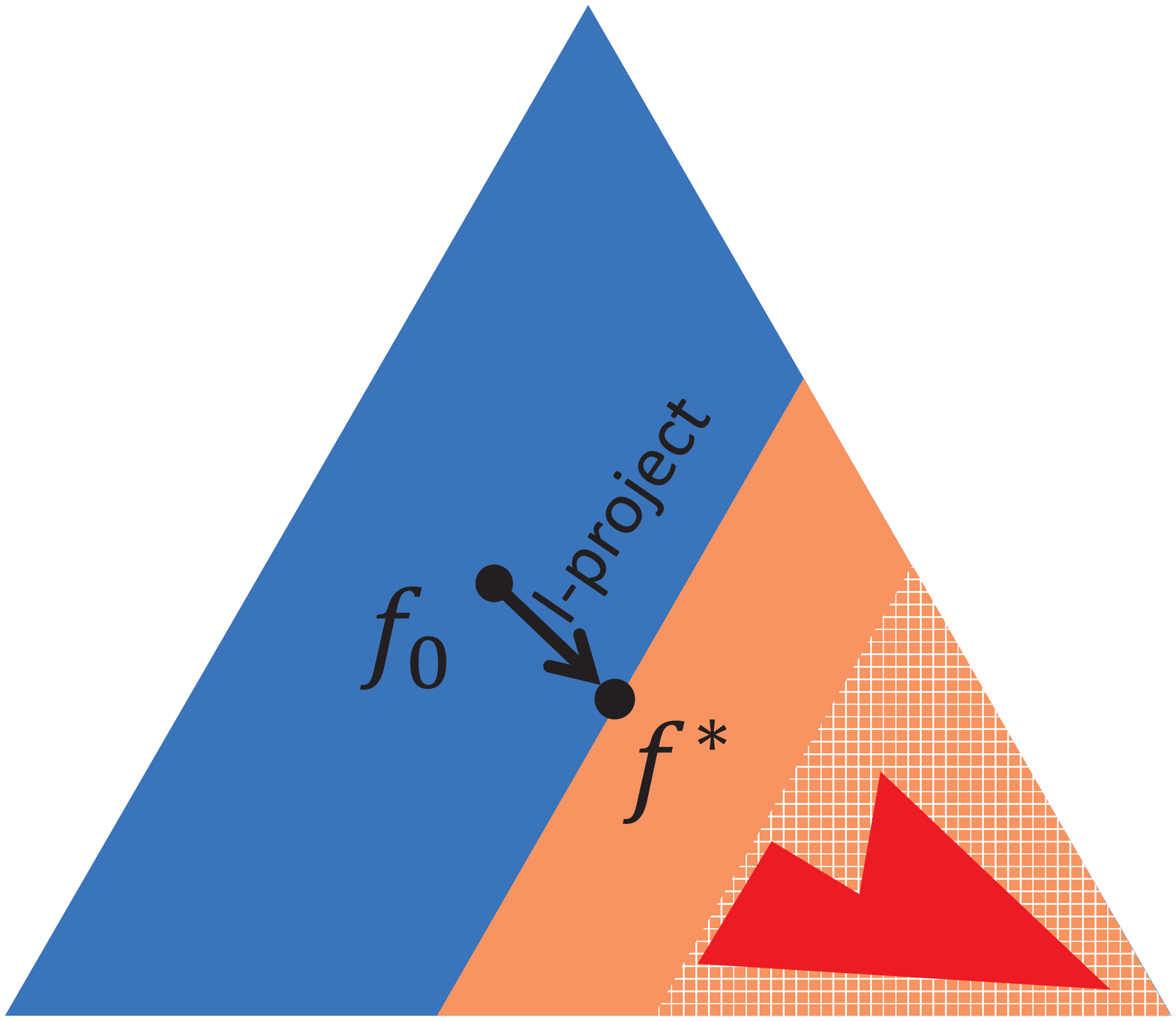}
\label{figure: simplex_min_distance}}
\hfil
\subfloat[A KL ball of radius \(c^D\) around \(f^*\) is excluded from \(\left\{q(f)\geq c^S\right\}\) in the new \(\Gamma_1\)]{\includegraphics[trim=160 0 230 0, clip,width=0.3\textwidth]{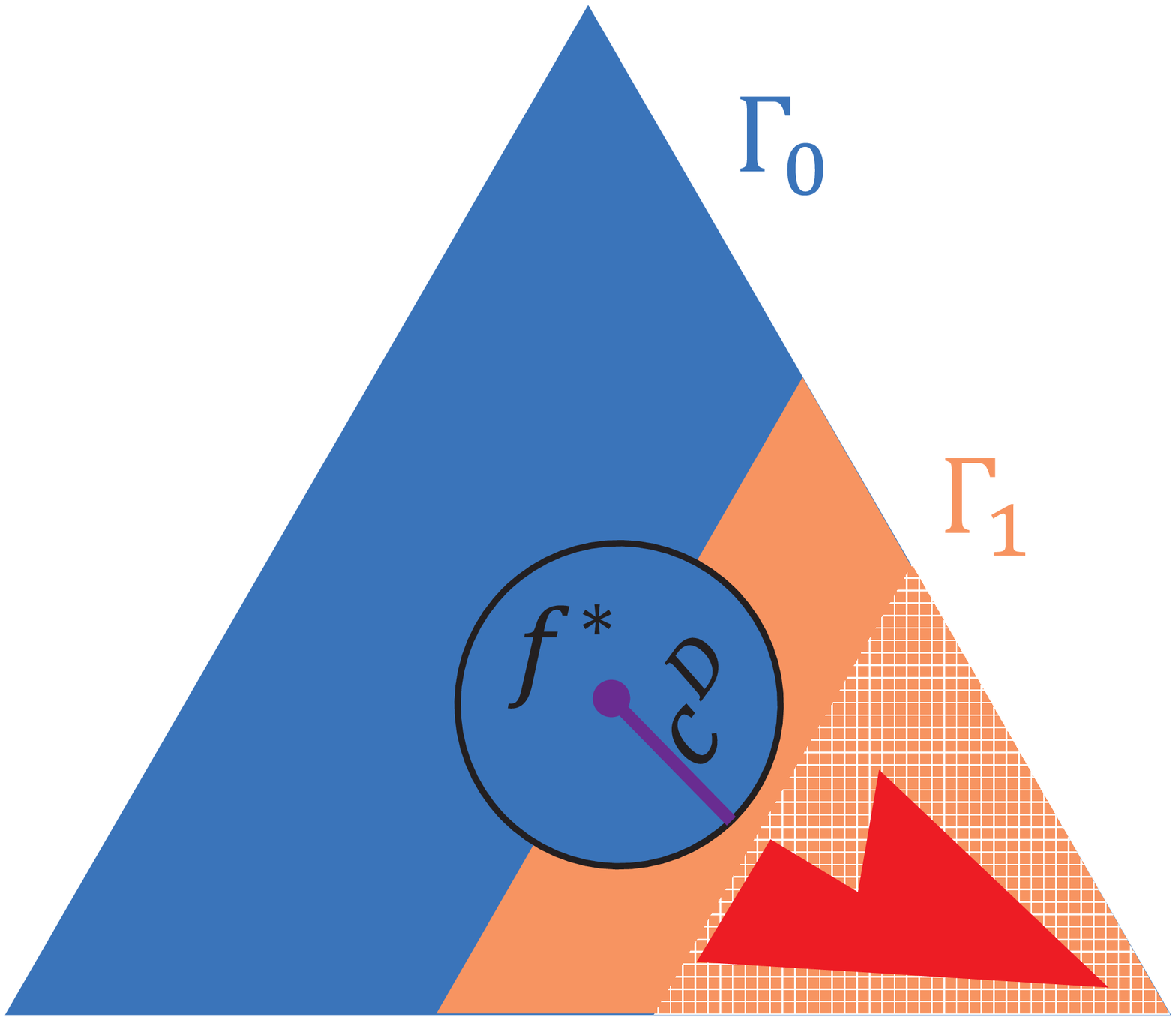}
\label{figure: simplex_new_algorithm}}
\caption{The probability simplex with \(m=3\) for our setup is illustrated. Each corner corresponds to a different deterministic distribution. The pre-change distribution \(f_0\) and its I-projection \(f^*\) onto \(\left\{q(f)\geq c^S\right\}\) are shown. Decision regions (old and new) \(\Gamma_0\) and \(\Gamma_1\) are shown in blue and light orange respectively. The (nonconvex) set of post-change distributions \(\mathcal{P}_1\) is shown in red.}
\label{figure: system_and_scheme}
\end{figure*}

First, we consider the composite hypothesis testing problem. Given \(n\), the null hypothesis, \(\mathbf{H}_0\), is true if \(X_1^n\) are i.i.d. \(f_0\) and the alternative hypothesis, \(\mathbf{H}_1\), is true if i.i.d. \(f_1\). Our goal is to minimize the maximum probability of misdetection given an upper bound on the probability of false alarm. An example decision rule region for the problem is given in Fig. \ref{figure: simplex} for \(m=3\). Once we define the decision regions \(\Gamma_0=\left\{\hat{\mathbf{H}}=\mathbf{H}_0\right\}\) and \(\Gamma_1=\mathcal{P}^n\setminus\Gamma_0=\left\{\hat{\mathbf{H}}=\mathbf{H}_1\right\}\) the problem can be stated as follows.
\begin{align}
    \inf_{\Gamma}\sup_{f_1}\quad &P_{f_1,1}\left(\hat {f}_{X_1^n}\in\Gamma_0\right)\\
    \text{subject to}\quad &P_\infty\left(\hat {f}_{X_1^n}\in\Gamma_1\right)\leq\alpha
\end{align}

Second, we consider the transient change detection problem \cite{shewhart} with unknown post-change distribution. The transient change detection problem minimizes the probability of misdetecting a single transient change within a specified window after the change. Without loss of generality (w.l.o.g.), we aasume the change in distribution is permanent but has to be detected in a bounded time \(n\). The maximization is subject to an upper bound on the probability of false alarm within any window of size \(n_\alpha\). If \(t_a\) denotes an arbitrary alarm (stopping) time of \(X_1^\infty\), the problem can be formulated as
\begin{align}
    \inf_{t_a}\quad &\sup_{f_1,t_1}P_{f_1,t_1}\left(t_a\geq t_1+n|t_a\geq t_1\right) \label{equation: tcdmd}\\
    \text{subject to}\quad &\sup_kP_\infty\left(k\leq t_a<k+n_\alpha\right)\leq\tilde{\alpha} \label{equation: tcdfa}
\end{align}
where we assume the worst post-change distribution. The optimal solution to this problem is unknown even for known post-change distribution except for the special case \(n=1\) subject to a minimum ARL \cite{shewhart}.

Finally, we consider the quickest change detection problem with unknown post-change distribution. The quickest change detection in Lorden's sense \cite{procedures} involves minimizing WADD given a minimum ARL. Assuming the worst post-change distribution, the problem can be formulated as
\begin{align}
    \inf_{t_a}\quad &\sup_{f_1,t_1}\esssup_{X_1^{t_1-1}} E_{f_1,t_1}\left(\left(t_a-t_1+1\right)^+\middle|X_1^{t_1-1}\right) \label{equation: lorden}\\
    \text{subject to}\quad &E_\infty t_a\geq\gamma. \label{equation: lorden2}
\end{align}
Optimal solution to this problem is also unknown. Therefore, we compare our solution of \eqref{equation: tcdmd}, \eqref{equation: tcdfa} and \eqref{equation: lorden}, \eqref{equation: lorden2} with frequently used (asymptotically optimal) methods in the literature.

\section{Motivation} \label{section: motivation}
The first problem is an instance of composite hypothesis testing \cite{signal}, without a parametric model under \(\mathbf{H}_1\). A typical detector in this case will pick \(\mathbf{H}_1\) if the empirical distribution, \(\hat{f}=\hat{f}_{X_1^n}\) is closer to a candidate distribution \(f_1\) than it is to \(f_0\). Even in that case, however, it is not necessarily true that \(X_1^n\) is drawn from some \(f_1\). It may still be drawn from \(f_0\), yet the empirical distribution might look as if it is drawn from some \(f_1\), leading to a false positive. We call such strings that are drawn from \(f_0\), but satisfy \(\hat{f}\in\mathcal{P}_1\) an outlier.

For clarity we provide insight from large deviations theory where the main results are asymptotic, but our contributions apply to the non-asymptotic regime as well. Next, we study the way outliers occur, when they occur using Sanov's theorem. See \cite{stochastic} for a detailed treatment.
\begin{theorem*}[Sanov]
For any continuous and quasiconcave function \(q:\mathcal{P}\to\mathbb{R}\), p.m.f. \(f_0\) and constant \(c\neq\sup_f q(f)\),
\begin{align}
\lim_{n\to\infty} -\frac{1}{n}\log P_\infty\left(q\left(\hat{f}_{X_1^n}\right)\geq c\right)&= \inf_{q(f)\geq c}I\left(f\middle\|f_0\right)
\end{align}
\end{theorem*}
\noindent
The proof of the most general form can be found in \cite{large}. Note that the interesting case is when \(q(f_0)<c<\sup_f q(f)\) and that we have the equality in Sanov's theorem, since \(\left\{q(f)\geq c\right\}\) is the closure of its interior as shown in Fig. \ref{figure: simplex_min_distance}, where we also illustrate Sanov's theorem. The solution of the constrained convex optimization \(\argmin_{q(f)\geq c} I\left(f\middle\|f_0\right)\) is called the information projection or I-projection of \(f_0\) onto the superlevel set \(\left\{q(f)\geq c\right\}\) and is denoted by \(f^*\). The theorem gives an asymptotic result in the number of samples and provides the exact characterization of the exponent at which the probability of an outlier decays. This implies that the most likely way for a sustained deviation to occur is the empirical distribution of the associated string to look as if it is drawn from the I-projection \(f^*\). The probability of this particular way of deviation dominates the probability of all others. An approximate version of this result also holds for a finite string of length \(n\) if Stirling's approximation is satisfied. In the next section, we give our algorithm which is based on using the projected distribution as the most likely outlier distribution. Subsequently, we evaluate the detection performance in the non-asymptotic region to justify its applicability.

\section{Information Projection Test} \label{section: algorithm}
The information projection test (IPT) exploits the observation that outliers occur in a particular way with high probability, while there is no such structure for the deviations. We describe the fixed window size IPT for hypothesis testing and transient change detection here and the variable size IPT for the quickest change detection in Subsection \ref{subsection: qcd}. Given a window size \(n\), the fixed window IPT involves the following steps.
\begin{enumerate}
    \item Pick \(c^S,c^D\geq 0\) and window size \(n\).
    \item Find the I-projection \(f^*=\argmin_{q(f)\geq c^S}I\left(f\middle\|f_0\right)\).
    \item At time \(k\geq n\), find \(S_k=q\left(\hat{f}_{X_{k-n+1}^k}\right)\). \label{enumerate: return}
    \item If \(S_k\geq c^S\), find \(D_k=I\left(\hat{f}_{X_{k-n+1}^k}\middle\|f^*\right)\); else, return to Step \ref{enumerate: return}.
    \item If \(D_k\geq c^D\), stop and claim a change has happened; else, return top Step \ref{enumerate: return}. \label{enumerate: stop}
\end{enumerate}
Note that Step \ref{enumerate: stop} determines if the empirical distribution \(\hat{f}\) is close to the most likely fistribution for an outlier, \(f^*\). This process is illustrated in Fig. \ref{figure: simplex_new_algorithm}. In the rest of the paper, we call the relative entropy \(I\left(f\middle\|f^*\right)\) also as the relative log likelihood function (RLLF) with respect to \(f^*\), referring to the comparison of the likelihood of the projection distribution with that of the empirical distribution of the observation. We also use \(c\) to denote \(\left(c^S,c^D\right)\).

\section{Performance Guarantees} \label{section: bounds}
\subsection{Composite Hypothesis Testing}
The next two theorems bound the probability of false alarm and misdetection as a function of \(c\). We show that the false positive probability decays exponentially with \(c^D+\frac{\left(\left(c^S-q_0\right)^+\right)^2}{2L^2}\) and that the worst case probability of misdetection decays with rate \(\left(\frac{\underline{q}-q_0-\left(c^S-q_0\right)^+}{\sqrt{2}L}-\sqrt{c^D}\right)^2\).
\begin{theorem} \label{theorem: false-alarm}
For any \(c\) such that \(c^S<\sup_fq(f)\),
\begin{align}
    P_\infty\left(\hat{f}_{X_1^n}\in\Gamma_1\right)&\leq \left(n+1\right)^m\exp\left(-n\left(c^D+\frac{\left(\left(c^S-q_0\right)^+\right)^2}{2L^2}\right)\right).
\end{align}
\end{theorem}

\begin{theorem} \label{theorem: misdetection}
For any \(c\) such that \(\left(c^S-q_0\right)^++\sqrt{2L^2c^D}<\underline{q}-q_0\),
\begin{align}
    \sup_{f_1}P_{f_1,1}\left(\hat{f}_{X_1^n}\in\Gamma_0\right)&\leq \left(n+1\right)^m\exp\left(-n\left(\frac{\underline{q}-q_0-\left(c^S-q_0\right)^+}{\sqrt{2}L}-\sqrt{c^D}\right)^2\right).
\end{align}
\end{theorem}

\subsection{Transient Change Detection}
In this subsection, we prove corollaries to Theorem \ref{theorem: false-alarm} and \ref{theorem: misdetection} that provide a bound on the transient change detection performance of IPT under the assumption of unknown post-change distribution. The transient change detection is regards the pair of metrics: the worst case probability of false alarm and worst case probability of misdetection within a given length of interval. Since we are given a window of \(n\) samples to detect the change and \(n\ll n_\alpha\), we utilize a fixed window IPT. This problem was first proposed in \cite{tcdfirst} and is also studied recently in \cite{shewhart,performance}.

\begin{corollary} \label{corollary: tcd}
For any \(c\) such that \(\left(c^S-q_0\right)^++\sqrt{2L^2c^D}<\underline{q}-q_0\) the IPT stopping time \(t_I\) with fixed window length \(\frac{n+1}{2}\) satisfies
\begin{align}
    \sup_kP_\infty\left(k\leq t_I<k+n_\alpha\right)&\leq 1-\left(1- \exp(-\nu n)\right)^{\left\lceil\frac{2n_\alpha}{n+1}\right\rceil}\\
    \sup_{f_1,t_1}P_{f_1,t_1}\left(t_I\geq t_1+n\middle|t_I\geq t_1\right)&\leq \exp\left(-\eta n\right)
\end{align}
where
\begin{align}
    \nu&= \frac{c^D}{2}+\frac{\left(\left(c^S-q_0\right)^+\right)^2}{4L^2}-\frac{m\log\left(\frac{n+3}{2}\right)}{n+1}\\
    \eta&= \left(\frac{\underline{q}-q_0-\left(c^S-q_0\right)^+}{2L}-\sqrt{\frac{c^D}{2}}\right)^2-\frac{m\log\left(\frac{n+3}{2}\right)}{n+1}.
\end{align}
\end{corollary}

\subsection{Quickest Change Detection} \label{subsection: qcd}
In this subsection, we apply IPT to the quickest change detection problem with an unknown post-change distribution. To minimize the average detection delay under the worst change time \(t_1\) and pre-change realizations \(X_1^{t_1-1}\), we modify IPT to have an effective window that only contains the samples with a positive drift from the minimum the random walk has visited, similar to the CUSUM algorithm. We also use a restart mechanism if the empirical distribution is similar to a most likely false alarm, similar to the IPT with fixed window size. Since, at any given time \(k\), the effective window size \(n_k\) is varying, we determine define the most likely deviation as a function of the window size, \(f_n^*\).
\begin{align}
    f_n^*&= \argmin_{q(f)\geq\frac{c^S}{n}}I\left(f\middle\|f_0\right)
\end{align}
For any \(k\), \(\tau_k\) denotes the last time the IPT has been restarted since the empirical distribution resembles a false alarm. So, \(\tau_1=1\) and for any \(k\),
\begin{align}
    (S_k,i_k)&= \max_{\tau_k\leq i\leq k+1}(k-i+1)q\left(\hat{f}_{X_i^k}\right)
\end{align}
We the index \(i_k\) that maximizes \(S_k\) to define the effective window size \(n_k=k-i_k+1\). Finally, we compare the proximity of the empirical distribution over the effective window with respect to the most likely false alarm given the window size \(n_k\). If \(D_k\) is large, we stop the observation process and raise an alarm, else we conclude a false alarm and reset the algorithm. The threshold for this step is allowed to vary with \(n\), so we select a sequence of \(c_n^D\geq 0\) instead of a constant \(c^D\).
\begin{align}
    D_k&= I\left(\hat{f}_{X_{i_k}^k}\middle\|f_{n_k}^*\right)\\
    \tau_{k+1}&= \begin{cases}
    k+1, &S_k\geq c^S, D_k<c_{n_k}^D\\
    \tau_k, &\text{otherwise} 
    \end{cases}\\
    t_I&= \inf\left\{k\middle|\left(S_k,D_k\right)\geq\left(c^S,c_{n_k}^D\right)\right\}
\end{align}
We also use shorthand notations \(c_n\) and \(Q(i,j)\) to refer to \(\left(\frac{c^S}{n},c_n^D\right)\) and \((j-i+1)q\left(\hat{f}_{X_i^j}\right)\) respectively. We define the auxiliary stopping time \(t_S=\inf\left\{k\middle|S_k\geq c^S\right\}\) where only the first statistic crosses the threshold and we define \(t_S^k\) and \(t_I^k\) to be the stopping times starting at \(k\), counting only the observations after \(X_{k-1}\).

In the following theorem we show that the ARL of our method can be lower bounded in terms of \(c\). We prove the theorem where \(q\) is the expectation operator but the general result can be obtained by constructing a separating hyperplane between the convex and closed set \(\mathcal{P}_1\) and the initial distribution \(f_0\notin\mathcal{P}_1\). Then, step 1 of the IPT holds only if the hyperplane is crossed. Thus, a lower bound on \(E_\infty t_S\) can be formulated using this halfspace instead of \(\mathcal{P}_1\).
\begin{theorem} \label{theorem: arl}
If \(c_n^D=c^D\geq \frac{2\left|q_0\right|\underline{q}}{(1+\rho)L^2}\) for \(n>(1+\rho)\frac{c^S}{\underline{q}}\) for some \(\rho>0\), then, the ARL of IPT is bounded as
\begin{align}
    E_\infty t_I&\geq \frac{\exp(v^*c^S)}{2^m\exp\left(-\frac{2\left|q_0\right|}{L^2}c^S\right)+\frac{\left((1+\rho)\frac{c^S}{\underline{q}}+1\right)^m\exp\left(-(1+\rho)\frac{c^D}{\underline{q}}c^S\right)}{1-\left((1+\rho)\frac{c^S}{\underline{q}}+1\right)^{\frac{m\underline{q}}{(1+\rho)c^S}}\exp\left(-c^D\right)}}\to\exp\left(\left(v^*+\frac{2\left|q_0\right|}{L^2}\right)c^S\right)
\end{align}
where \(\psi(v)= E_\infty\exp(vX)\) and \(v^*>0\) satisfies \(\psi(v^*)=1\).
\end{theorem}
The following lemma proves a bound on the stopping time for any change time \(t_1\) and observations \(X_1^{t_1-1}\).
\begin{lemma} \label{lemma: stopping}
For any \(t_1,f_1\) and \(X_1^{t_1-1}\), the IPT stopping time \(t_I\) is bounded by
\begin{align}
    t_I&\leq t_S^{(j)}+t_I^{t_S^{(j)}+1} \label{equation: stopping}
\end{align}
for any \(j=0,\dots\) where \(t_S^{(j)}\) is the \(j\)th time \(S_k\) crosses \(c^S\), i.e. \(t_S^{(0)}=0\) and \(t_S^{(j+1)}=t_S^{(j)}+t_S^{t_S^{(j)}+1}\). Further, for any \(j=1,\dots\) and \(t_S^{(j-1)}<k\leq t_S^{(j)}\),
\begin{align}
    t_S^{(j)}& \leq k+t_S^k-1. \label{equation: auxilary stopping}
\end{align}
Thus, for \(j^*=\min\left\{j\middle|t_S^{(j)}\geq t_1\right\}\),
\begin{align}
    t_I&\leq t_1+t_S^{t_1}-1+t_I^{t_S^{(j^*)}+1}.
\end{align}
\end{lemma}
In the next lemma, we prove that the average delay is bounded by a function of the average delay when \(t_1=1\) using the bound we have found in Lemma \ref{lemma: stopping}. This lemma shows that the worst average delay IPT experiences is bounded by the average delay up to a multiplicative constant.
\begin{lemma} \label{lemma: wadd}
For any \(f_1,t_1\) and \(X_1^{t_1-1}\), the average delay is bounded as follows.
\begin{align}
    E_{f_1,t_1}\left(\left(t_I-t_1+1\right)^+\middle|X_1^{t_1-1}\right)&\leq E_{f_1,1}\left(t_S+t_I\right)
\end{align}
\end{lemma}
Before showing a bound on the WADD, we bound the average detection delay when change happens at \(t_1=1\) asymptotically as \(c^S\to\infty\).
\begin{lemma} \label{lemma: add}
For any \(f_1\), if \(c_n^D=0\) for \(n\leq(1+\rho)\frac{c^S}{\underline{q}}\) for some \(\rho>0\), then, as \(c^S\to\infty\)
\begin{align}
    E_{f_1,1}t_I&\leq \frac{c^S}{q_1}.
\end{align}
\end{lemma}
The next theorem is the asymptotic characterization of the WADD of IPT. In the proof we first derive a non-asymptotic bound and then consider the limit as \(c^S\to\infty\).
\begin{theorem} \label{theorem: wadd}
If \(c_n^D=0\) for \(n\leq(1+\rho)\frac{c^S}{\underline{q}}\) for some \(\rho>0\), then, as \(c^S\to\infty\),
\begin{align}
    WADD(t_I)&\leq \frac{2c^S}{\underline{q}}
\end{align}
\end{theorem}
Lorden showed in his seminal work \cite{procedures} that the optimal WADD for any change detection algorithm with ARL at least \(\gamma\) is \(\frac{\log\gamma}{I\left(f_1\middle\|f_0\right)}\) asymptotically as \(\gamma\to\infty\) even if the post-change distribution \(f_1\) is known. Using Theorem \ref{theorem: arl} and \ref{theorem: wadd}, we prove that IPT is asymptotically optimal in Lorden's criteria up to a multiplicative constant.
\begin{corollary}
For \(\rho>0, c^D\geq\frac{2\left|q_0\right|\underline{q}}{(1+\rho)L^2}\) and \(c_n^D=\begin{cases}0,& n\leq(1+\rho)\frac{c^S}{\underline{q}}\\c^D,& \text{otherwise}\end{cases}\), as \(\gamma\to\infty\),
\begin{align}
    WADD(t_I)=\frac{\log\gamma}{\underline{q}\left(\frac{v^*}{2}+\frac{\left|q_0\right|}{L^2}\right)}.
\end{align}
\end{corollary}
\begin{IEEEproof}
Use Theorem \ref{theorem: arl} and \ref{theorem: wadd}.
\end{IEEEproof}

We end our discussion on the quickest change detection problem by describing two special cases.
\subsubsection{Special Case: Change in Mean}
For the special case where the change is in the mean, IPT reduces to a CUSUM algorithm with restart.
\begin{align}
    q\left(\hat{f}_{X_i^k}\right)&= \frac{1}{k-i+1}\sum_{j=i}^kX_j\\
    S_k&= \max_{\tau_k\leq i\leq k}\sum_{j=i}^kX_j\\
    &= \begin{cases}
    \left(S_{k-1}+X_k\right)^+, &\tau_k\neq k\\
    \left(X_k\right)^+, &\tau_k=k
    \end{cases}
\end{align}
Further, the most likely deviations \(f_n^*\) become the exponentially tilted distributions with mean \(\max\left\{\frac{c^S}{n},q_0\right\}\).
\begin{align}
    f_n^*&= \argmin_{\sum_aaf(a)\geq\frac{c^S}{n}}I\left(f\middle\|f_0\right)\\
    f_n^*(a)&= f_0(a)\exp\left(ra-\Lambda\left(r\right)\right)\\
    D_k&= I\left(\hat{f}\middle\|f_{n_k}^*\right)\\
    &= I\left(\hat{f}\middle\|f_0\right)+\Lambda(r)-r\sum_aa\hat{f}(a)
\end{align}
where
\begin{align}
    \Lambda(r)&= \log E_\infty\exp\left(rX\right)\\
    E_\infty\left(\exp\left(rX-\Lambda\left(r\right)\right)X\right)&= \max\left\{\frac{c^S}{n},q_0\right\}.
\end{align}

\subsubsection{Special Case: Existence of a Representative \(f_1\in\mathcal{P}_1\) for Log-Likelihood Ratio}
For the special case where there exists \(f_1\in\mathcal{P}_1\) and \(\rho>0\) such that for all \(f\in\mathcal{P}_1\),
\begin{align}
    \norm{f-f_1}_1&\leq (1-\rho)\frac{I\left(f_1\middle\|f_0\right)}{\max_a\log\frac{f_1(a)}{f_0(a)}}
\end{align}
the function \(q\left(f\right)=I\left(f\middle\|f_0\right)-I\left(f\middle\|f_1\right)\) is a quasiconcave and Lipschitz continuous function with \(q_0=q\left(f_0\right)=-I\left(f_0\middle\|f_1\right)<0\) and for any \(f\in\mathcal{P}_1\),
\begin{align}
    q(f)&= I\left(f\middle\|f_0\right)-I\left(f\middle\|f_1\right)\\
    &= \sum_af(a)\log\frac{f_1(a)}{f_0(a)}\\
    &= \sum_a\left(f(a)-f_1(a)+f_1(a)\right)\log\frac{f_1(a)}{f_0(a)}\\
    &= I\left(f_1\middle\|f_0\right)+\sum_a\left(f(a)-f_1(a)\right)\log\frac{f_1(a)}{f_0(a)}\\
    &\geq I\left(f_1\middle\|f_0\right)-\norm{f-f_1}_1\max_a\log\frac{f_1(a)}{f_0(a)}\\
    &\geq \rho I\left(f_1\middle\|f_0\right).
\end{align}
Thus, for all \(f\in\mathcal{P}_1\), \(0<\rho I\left(f_1\middle\|f_0\right)=\underline{q}\leq q(f)\). Further, we can interpret \(Q(1,k)=kq\left(\hat{f}_{X_1^k}\right)\) as a random walk with i.i.d. steps of the log-likelihood ratio \(\log\frac{f_1\left(X_i\right)}{f_0\left(X_i\right)}\). Finally, \(S_k=\max_{\tau_k\leq i\leq k}\sum_{j=i}^k\log\frac{f_1\left(X_i\right)}{f_0\left(X_i\right)}\) becomes the CUSUM algorithm with restart.

\section{Comparison to Other Detection Schemes} \label{section: comparison}
We compare IPT with various change detection algorithms proposed in the literature in terms of ROC and Lorden's criteria.

\begin{figure*}[!t]
\centering
\begingroup
\captionsetup[subfigure]{width=0.225\textwidth}
\subfloat[Composite hypothesis testing problem ROC, \(m=3\)]{\includegraphics[width=0.24\textwidth]{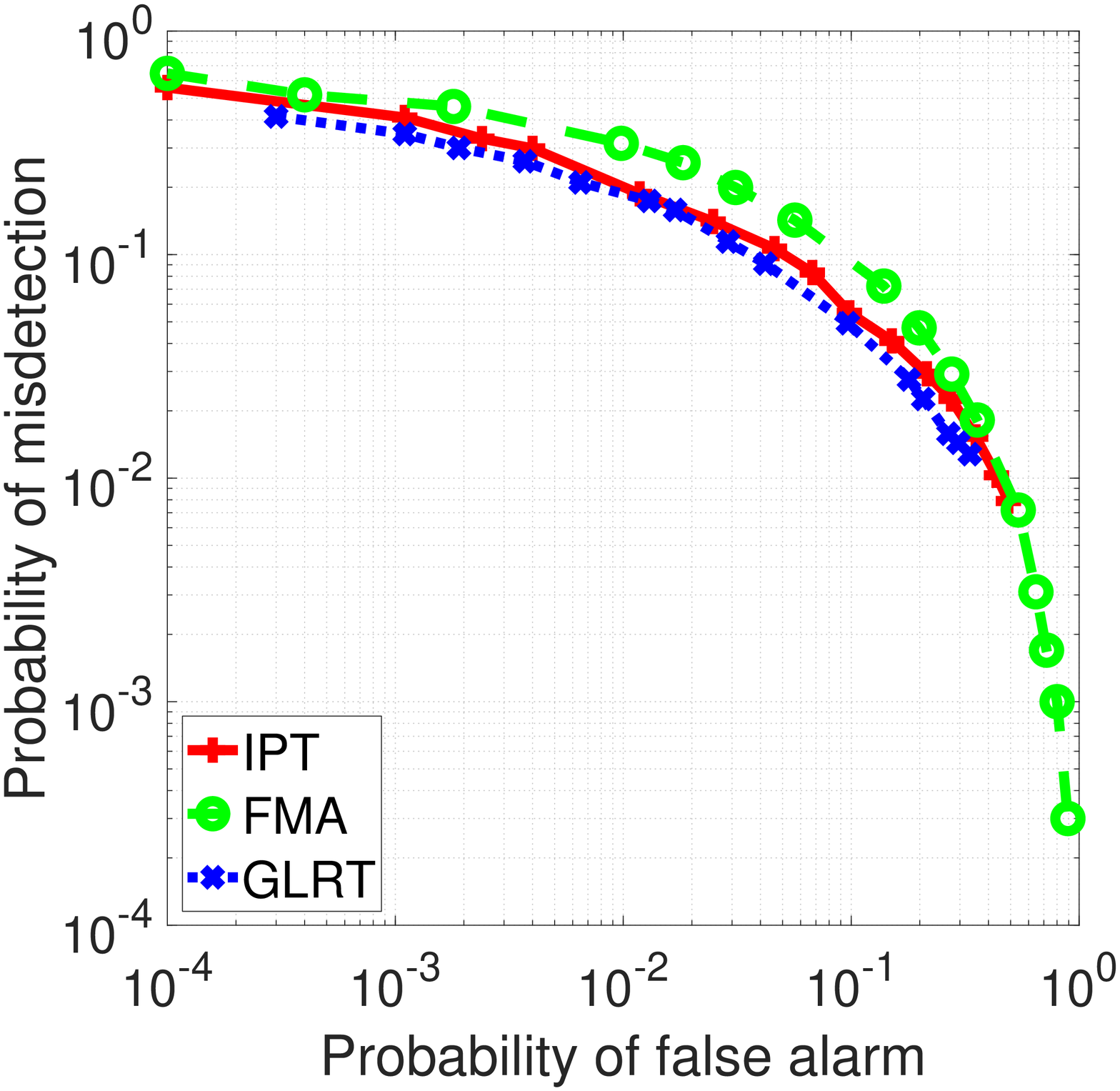}
\label{figure: ROC}}
\hfil
\subfloat[Transient change detection problem ROC for a discrete Gaussian alphabet, \(m=11\)]{\includegraphics[width=0.24\textwidth]{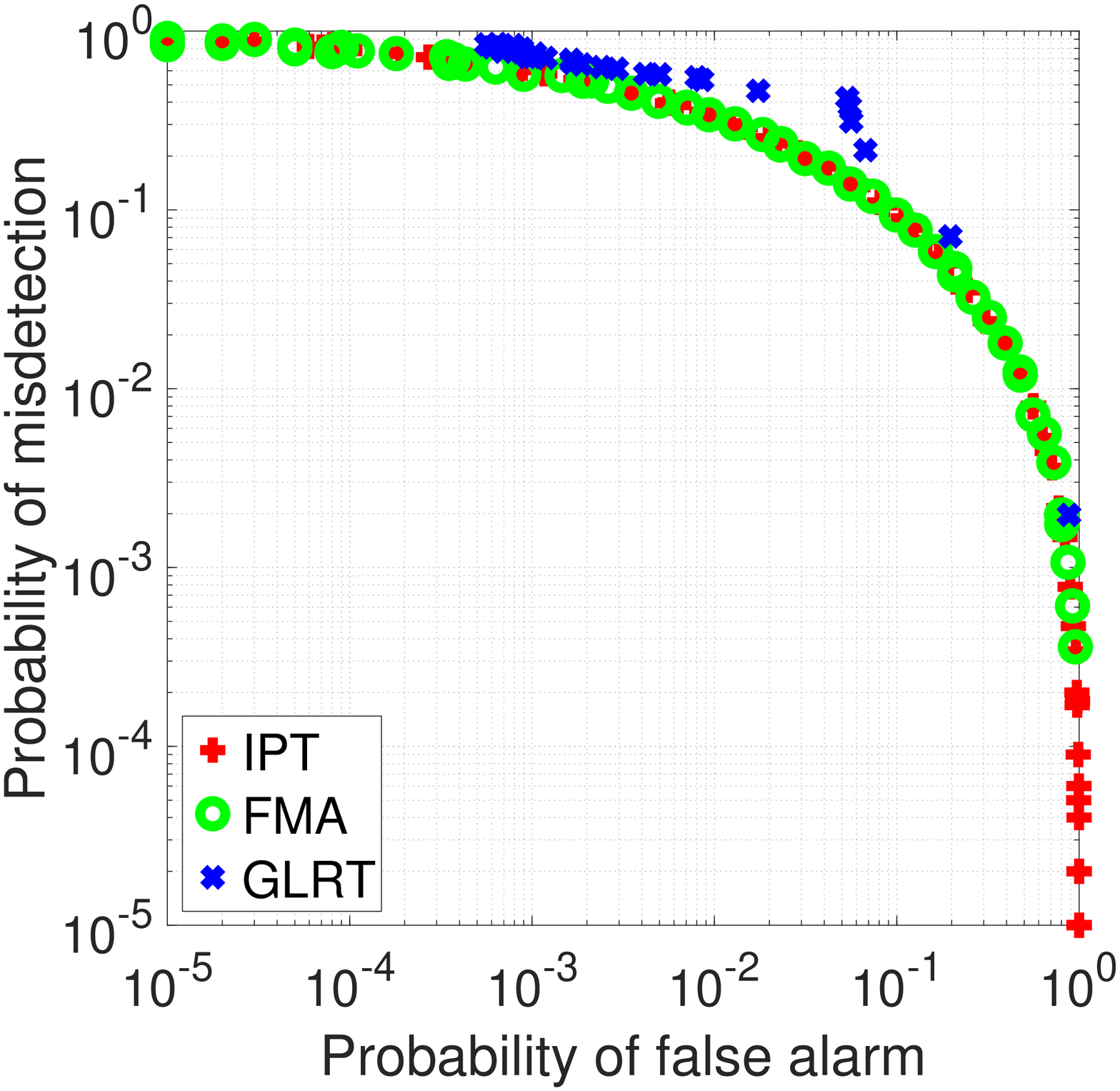}
\label{figure: TCD}}
\hfil
\subfloat[Quickest change detection problem ARL vs WADD, \(m=3\)]{\includegraphics[width=0.24\textwidth]{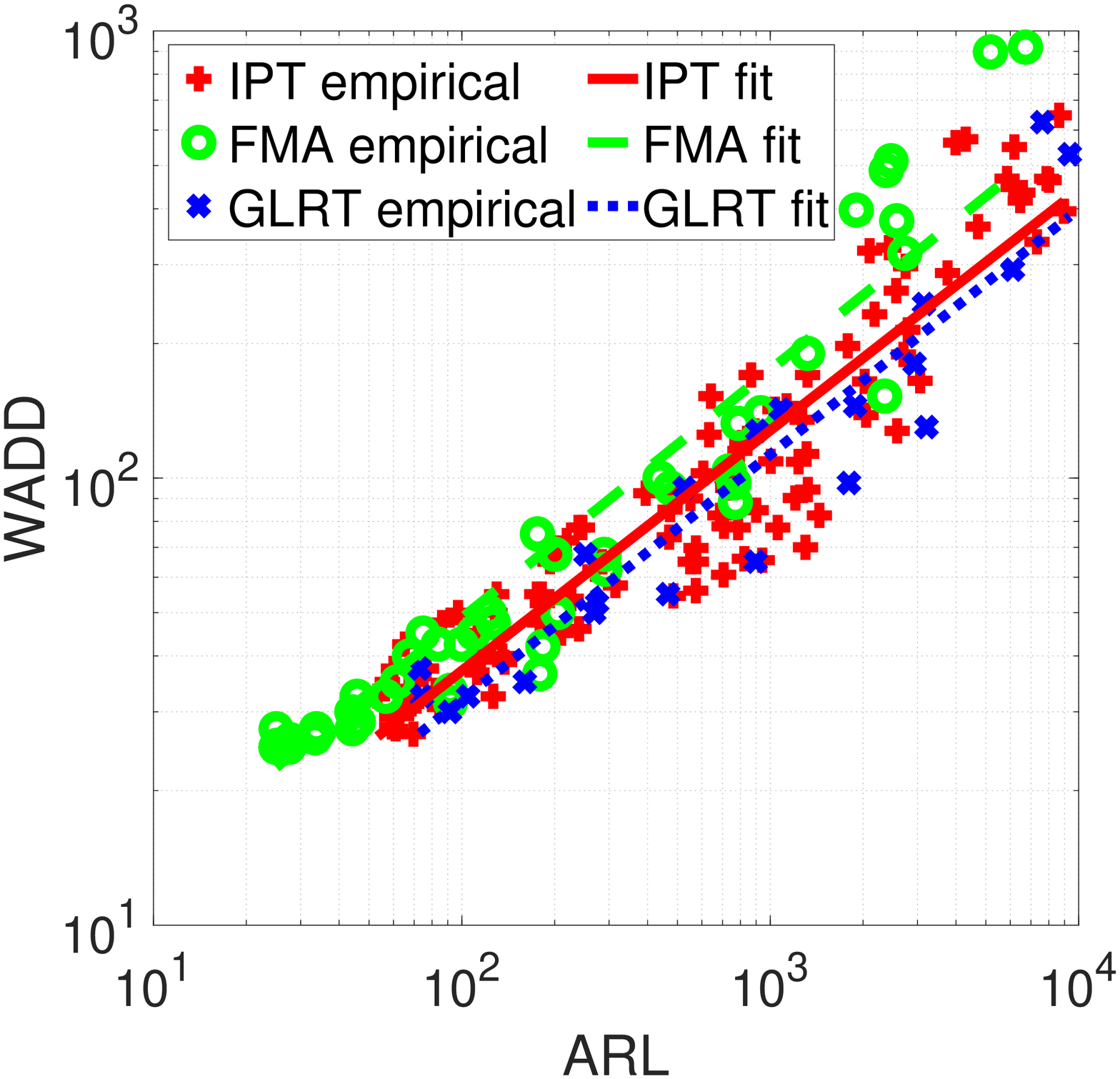}
\label{figure: delay}}
\hfil
\subfloat[Average time required to perform the test, \(m\propto n\)]{\includegraphics[width=0.24\textwidth]{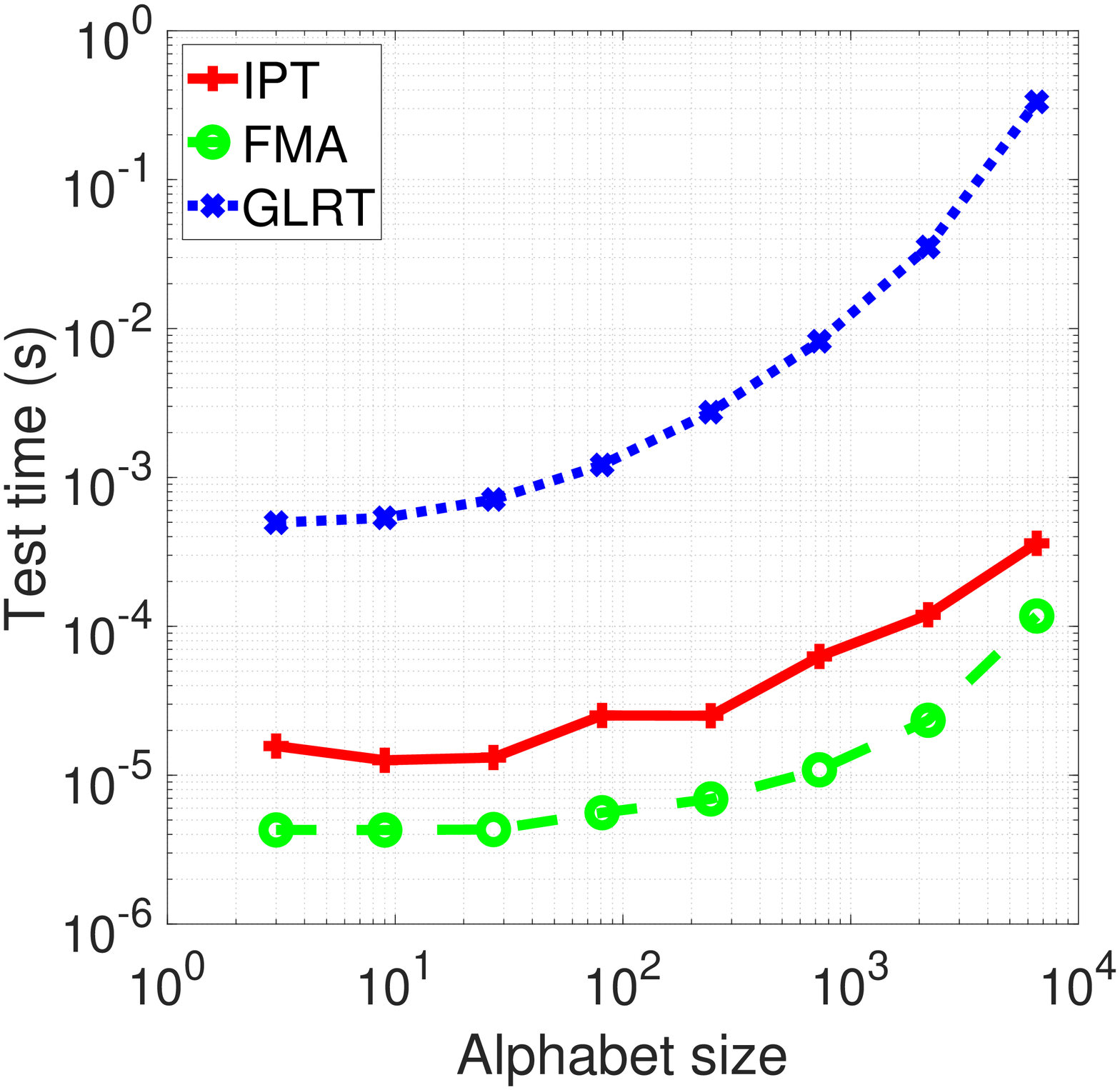}
\label{figure: time-complexity}}
\endgroup
\caption{IPT, FMA and GLRT are compared in terms of ROCs, Lorden's criteria and computational complexity.}
\label{figure: tests}
\end{figure*}

\subsection{Composite Hypothesis Testing} \label{subsection: cht}
For the hypothesis testing and transient change detection problems, we describe the simulation procedure and plot the empirical ROC, worst case probability of misdetection versus probability of false alarm, for each of the methods considered.

The ROC associated with our method is illustrated in Fig. \ref{figure: ROC} and compared with the ROC of an FMA test and GLRT. We use a ternary alphabet \(\{-1,0,1\}\), define the pre-change distribution to be uniform and the change in mean to satisfy \(\underline{q}=0.25\). The deviations are chosen to have \(n=25\). At this stage, we determine the I-projection of \(f_0\) on \(\mathcal{P}_1\). To find the optimum performance of our algorithm, we change \(c^D\) over \([2^{-8},2^{-3}]\). The curves in Fig. \ref{figure: ROC} show that, our proposed method is able to decrease the false positive rate compared to CUSUM-like FMA filter for the same misdetection rate and performs similar to the GLRT. IPT's average misdetection performance is within \(27\%\) of that of GLRT but performs \(34\%\) better than FMA over the measured region of the ROC. The area under the misdetection-false alarm curves are \(0.343,0.224\) and \(0.184\) for FMA, IPT and GLRT respectively.

\subsection{Transient Change Detection}
We consider the transient change detection problem under a family of discrete Gaussian distributions. Discrete Gaussian sampling is of particular interest in lattice-based cryptography \cite{crypto} and thereby in quantum-resilient cryptography \cite{quantum}. For hardware based solutions proposed to generate fast and true random numbers \cite{hardware}, detecting a change in the distribution of generated discrete Gaussian samples could increase the reliability against failures. Here, we only consider a finite alphabet discrete Gaussian with \(\mathcal{A}=\{-5,\dots,5\}, m=11, n=80, d_\alpha=200\) and \(f(k)=N\exp\left(-\frac{k^2}{2d^2}\right)\). The known pre-change distribution has \(\sigma_0^2=1\) and the post-change distribution satisfies \(\sigma_1^2\geq\underline{\sigma}^2=2\). For \(q(f)=\sigma_f^2\), the set of post-change distributions \(\mathcal{P}_1= q^{-1}([\underline{\sigma}^2,\infty))\) is a convex subset of \(\mathcal{P}\), thus we can utilize the IPT. For each method, we use a rolling window of size \(w=20<\frac{n+1}{2}\). The worst change point \(t_1\) is chosen among \(t_1=1,\dots,w\) and we randomly specify \(100\) different post-change distributions from \(\mathcal{P}_1\). Each data point in Fig. \ref{figure: TCD} is the empirical average of \(10^5\) worst case false alarms and worst case misdetection tests. With \(\frac{20}{11}\sim 2\) samples per letter, GLRT overfits the empirical data while estimating \(\sigma_1^2\) and performs worse than IPT and FMA.

\subsection{Quickest Change Detection}
For the quickest change detection problem, we test the IPT as described in Subsection \ref{subsection: qcd} in terms of WADD versus ARL under the test procedure described. Then, the empirical delay curves of different algorithms are compared.

In Fig. \ref{figure: delay}, we consider the quickest change detection performance of IPT, FMA and GLRT. The minimum WADD is plotted against the ARL under the same setting as in Subsection \ref{subsection: cht}. IPT provides a tradeoff between the mean test and GLRT with slightly worse detection delay compared to the latter. But in a scenario where computational costs would not allow a user to choose only according to the delay characteristics IPT may outperform both FMA and GLRT. A fast stream of data tracked for change at a central unit is such an example.

\section{Complexity} \label{section: complexity}
In this section we describe the complexity of the change detection algorithms: IPT, FMA and GLRT. We compare the computational complexity as a function of sample size \(n\), alphabet size \(m\) or error \(\epsilon\sim\norm{\tilde{f^*}-f^*}_1\) involved in I-projection for IPT and maximum likelihood estimation during GLRT. The FMA test using a rolling window has \(\Theta(n)\) complexity\footnote{\(\mathcal{O}(g)\) is the set of functions asymptotically upper bounded by \(g\) \cite{clrs}, i.e. \(\mathcal{O}(g)=\left\{f\middle|\exists d,u_0>0\ \text{such that}\ \forall u\geq u_0, 0\leq f(u)\leq dg(u)\right\}\) where, if \(u\) is a vector variable the inequalities are componentwise. Similarly, \(\Theta(g)=\left\{f\middle|f\in\mathcal{O}(g)\ \text{and}\ g\in\mathcal{O}(f)\right\}\).} to decide on the alarm time during execution. The GLRT forms a likelihood ratio for each new sample and therefore solves the following (nonconvex) optimization problem:
\begin{align}
    \min_{f_1\in\mathcal{P}_1}\quad &I\left(\hat{f}_{X_i^n}\middle\|f_1\right)
\end{align}
which (if \(\mathcal{P}_1\) is convex) is computationally equivalent to binary searching two Lagrange multipliers using an equation of \(m\) rational terms, thus it has \(\mathcal{O}\left(n+m\log\frac{1}{\epsilon}\log\log\frac{1}{\epsilon}\right)\) complexity. IPT omits this projection step by initially finding the most likely deviations and then comparing any realization to this distribution. Therefore, it has only \(\Theta(n+m)\) complexity. Since GLRT does not allow a recursive update, its complexity is the same for a sliding window method. Assuming the function \(q\left(\hat{f}\right)\) admits a recursive update in \(\Theta(1)\), IPT has overall \(\Theta(1)\) complexity for sliding windows. This is because \(I\left(\hat{f}\middle\|f^*\right)\) has a simple update rule. Similarly FMA with rolling window has \(\Theta(1)\) complexity. Thus, whereas IPT and FMA have linear complexity for rolling windows and constant complexity for sliding windows, GLRT has superlinear complexity in both cases.

For a data stream with packet rate \(R\) satisfying \(\frac{1}{R}\gg n+m\), GLRT can not respond to changes in the time series as fast as IPT or FMA. This is especially true if a worst case detection delay is used as a cost. With a rapidly increasing demand for social media content, video, distributed computation there will be a growing need for fast heuristics than for offline exact computation.

\begin{table}[ht]
\renewcommand{\arraystretch}{1.3}
\caption{Time complexity scaling}
\label{table: complexity}
    \centering
    \begin{tabular}{c|c|c|c|}
        \cline{2-4}
                                             & IPT & FMA & GLRT \\ \hline
        \multicolumn{1}{|c|}{init.} & \(\mathcal{O}\left(n+m\log\epsilon^{-1}\right)\) & \(\Theta(1)\) & \(\Theta(1)\) \\ \hline
        \multicolumn{1}{|c|}{rolling} & \(\Theta(d+m)\) & \(\Theta(d)\) & \(\mathcal{O}(n+m\log m\log\epsilon^{-1})\) \\ \hline
        \multicolumn{1}{|c|}{sliding} & \(\Theta(1)\) & \(\Theta(1)\) & \(\mathcal{O}(n+m\log m\log(d\epsilon)^{-1})\) \\ \hline
\end{tabular}
\end{table}

\subsubsection{Computational Complexity}
Using Ohio Supercomputer Center Owens cluster's single node of 128GB memory and 2.40GHz clock the test times varied as in Fig. \ref{figure: time-complexity} \cite{osc}. We increased the alphabet and sample sizes proportionally and averaged over the total number of tests. IPT performed \(25\) to \(670\) times faster than the GLRT where the gain from test time increased with the alphabet size.

\section{Applications to Economics and Climate Data} \label{section: applications}

In this section, we apply our outlier detection scheme, based on checking the RLLF of the most likely outlier distribution to the empirical distribution of the given period. We would like to test whether shifts of historic value are merely outliers or it happens due to an exogenous factor. To obtain the ground truth, we will use the long-term empirical distribution. Thus, the underlying assumption we make in the following data analysis is that, for \(r\) shifts of length \(n\) the overall cumulative duration of deviations is much smaller compared to the size of the data: \(rn \ll T\).
To analyze a period of interest, we pick a duration \(n\) and a threshold, \(c^S\), that matches the mean of the underlying data over that period.

\subsection{Daily Returns of Portfolios by Size} \label{subsection: market}
\begin{figure}[H]
    \centering
    \includegraphics[scale=0.2]{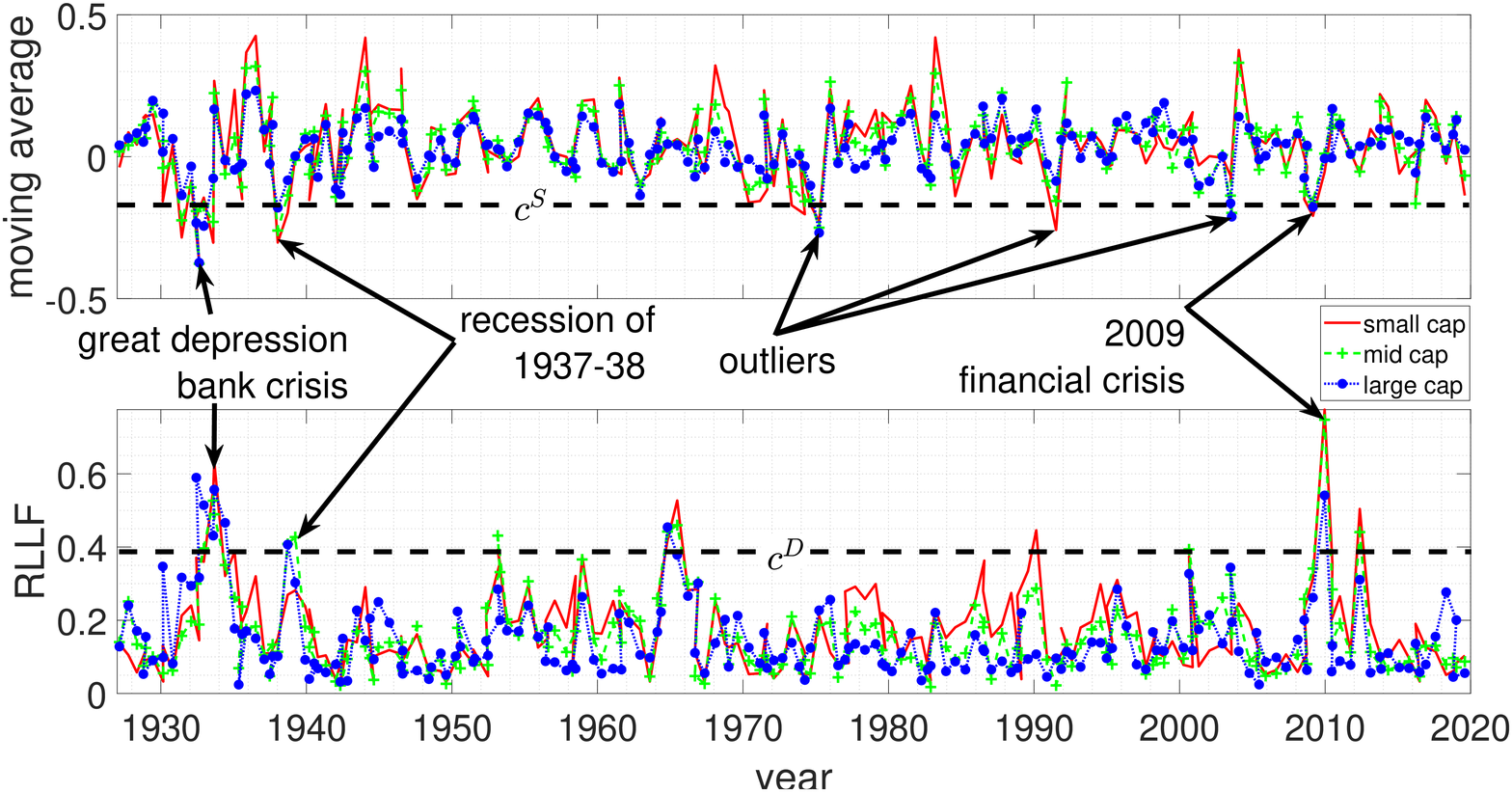}
    \caption{The moving average and the RLLF for the market indices with different market capitalization}
    \label{figure: industry_indices}
\end{figure}

We focus on portfolio returns with different market caps \cite{ken_french_data} to identify segments with a worse performance compared to the average behavior, sustained over a substantial duration of time. The primary question we answer is the following: ``is the identified negative sequence a rare event generated by the statistical nature of the market or is it driven by some exogenous factors with the potential to cause a financial crisis?'' The monthly averages of the return vary between \(0.8-0.9\). We quantized the daily percentage portfolio returns for different market caps over July 1926-March 2019 and selected a threshold \(c^S\) below the average returns. We then computed the I-projection of the long term empirical distribution over the distributions with mean less than the threshold. The \(n=6\) month moving average and the KL divergence of the observation window against the I-projection is given in Fig. \ref{figure: industry_indices}. With the right choice of \(n\) and \(c\) we were able to identify historically significant events like the great depression or the 2009 financial crisis. One of the most significant finding is that, even though there are other periods with mean return as low as these periods, the RLLF clearly differentiated these periods from the crises times, showing strength of our scheme.

\subsection{Oil Industry Market Index}
\begin{figure}[H]
    \centering
    \includegraphics[scale=0.2]{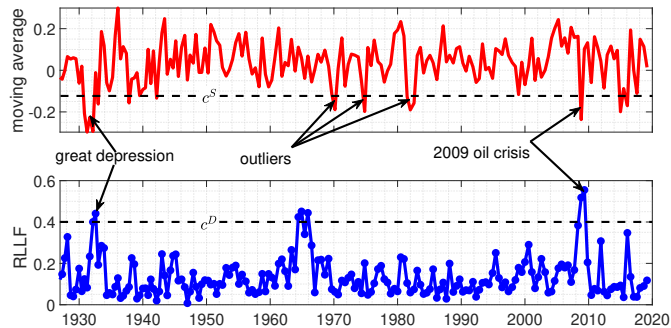}
    \caption{The moving average and the RLLF for the oil industry market index}
    \label{figure: oil_index}
\end{figure}

Similar to Subsection \ref{subsection: market}, this time we focus on market index for oil industry. We use monthly return series of Oil Industry Portfolio Index, obtained from Kenneth French Data Library \cite{ken_french_data} and covers the period from 1926-07-01 to 2019-03-31. The historical monthly return of the index is \(0.89\). We have run the experiment with \(c^S=-0.1\%\) and \(n=12\) months. In Fig. \ref{figure: oil_index}, we have \(7\) first threshold crosses, only \(2\) of which classified as change. The greatest divergence is for the \(2009\) oil crisis period. Unfortunately, the \(1973\) oil embargo is disregarded as a false alarm. This example shows the requirement of tuning for our method.

\subsection{Oxygen Isotope Data}
\begin{figure}[H]
    \centering
    \includegraphics[scale=0.2]{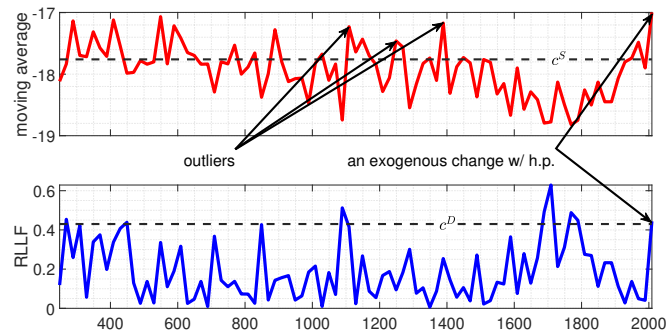}
    \caption{The moving average and RLLF for the oxygen isotopes and annual ice accumulation}
    \label{figure: climate}
\end{figure}

We analyze climate data collected from a polar cap that spans \(\sim 1800\) years to gain insight of our method and historic climate data \cite{thompson}. Fig. \ref{figure: climate} uses a moving average window of \(n=20\) years of quantized data of oxygen isotope density over the years \(226-2009\). We then select thresholds \(c^S\), half a standard deviation above the mean, and \(c^D\) and determine the I-projections. Note that the only point IPT classifies as change is the last 20-year period of above norm d18O levels. Although the threshold selections are not unique this intuitive approach with these hyperparameters gives insight that the last 20-year period is less likely to be an outlier rather than an effect of an exogenous change, ie. man made.

\section{Conclusion} \label{section: conclusion}
In a variety of applications, it is important to identify shifts from the typical behavior. However, not every shift from the norm marks a regime change or a deviation; instead it could be an outlier. In this paper, we developed a method that differentiates between outliers and change. To achieve that, we used results from large deviations theory, which gives us statistical characterization for the outliers. Our method tests a given string against the most likely way an outlier occurs and marks deviations using detection-theoretic tools.

Our method uses control parameters \(c^S\) and \(c^D\) to shape the decision region to determine the outliers. We have bounded the probability of false positive and misdetection as functions of these control parameters and shown that the detection delay is asymptotically optimal given an ARL.

We also emphasized IPT's computational simplicity in that it reduces GLRT's requirement to run non-convex and non-recursive optimization problems to solving convex optimization problems during initialization. This has proven to reduce the average running time per test 25 to 670 fold for varying alphabet sizes.

We also applied our algorithm to a variety of applications and drawed insights from the observed time series data. Our method verified the global warming phenomenon as highly unlikely to be an outlier, giving credence for it to be caused by factors (such as those man-made) leading to a statistical change in the indicator variables. Similarly, we tried to extract historically significant exogeneous events for the market within the data.

We have omitted proves that complicate the exposition whenever necessary but we believe similar results can be extended to the continuous alphabet case via binning.

Future work will be on proposing a metric that naturally and objectively combines sample delay for detection with computational complexity to address true delays as observed in high traffic data centers that detect changes in popularity, inappropriate content, fake accounts or attacks.

\appendices
\section{Inequalities}
The inequalities utilized throughout the paper are briefly provided in this subsection for easy reference.
\begin{theorem*}[Pythagorean Theorem for Relative Entropy \cite{elements}]
For any closed and convex set \(C\subset\mathcal{P}\), let \(f^*\) denote the I-projection of \(f_0\) on \(C\). Then, for any \(f\in C\),
\begin{align}
    I\left(f\middle\|f_0\right)&\geq I\left(f\middle\|f^*\right)+I\left(f^*\middle\|f_0\right).
\end{align}
\end{theorem*}
\begin{theorem*}[Pinsker's Inequality \cite{elements}]
For any pair \(f\) and \(f'\),
\begin{align}
    I\left(f\middle\|f'\right)&\geq \frac{\norm{f-f'}_1^2}{2}.
\end{align}
\end{theorem*}
\begin{corollary*}
For any \(f\) and \(f'\),
\begin{align}
    I\left(f\middle\|f'\right)\geq\frac{\norm{f-f'}_1^2}{2}\geq\frac{\left(q(f)-q(f')\right)^2}{2L^2}.
\end{align}
\end{corollary*}
\begin{IEEEproof}
The first inequality is the Pinsker's inequality. The second inequality is due to the Lipschitz property of \(q\).
\end{IEEEproof}
\begin{theorem*}[Inequality for the \(L_1\) Deviation \cite{inequalities}]
For any \(f_0\), \(n\) and \(\delta>0\),
\begin{align}
    P_\infty\left(\norm{\hat{f}_{X_1^n}-f_0}_1\geq\delta\right)&\leq 2^m\exp\left(-n\frac{\delta^2}{2}\right)
\end{align}
\end{theorem*}

\section{Proofs of Theorems}
\subsection{Proof of Theorem \ref{theorem: false-alarm}}
Given \(c\) such that \(c^S<\sup_fq(f)\), we have \(\Gamma_1= \left\{f\in\mathcal{P}^n\middle|q\left(f\right)\geq c^S,I\left(f\middle\|f^*\right)\geq c^D\right\}\) and
\begin{align*}
    P_\infty\left(\hat{f}_{X_1^n}\in\Gamma_1\right)&= \sum_{f\in \Gamma_1}P_\infty\left(\hat{f}_{X_1^n}=f\right)\\
    &\leq \sum_{f\in \Gamma_1}\exp\left(-nI\left(f\middle\|f_0\right)\right)\\
    &\leq \sum_{f\in \Gamma_1}\exp\left(-n\left(I\left(f\middle\|f^*\right)+I\left(f^*\middle\|f_0\right)\right)\right)\\
    &\leq \sum_{f\in \Gamma_1}\exp\left(-n\left(c^D+\frac{\left(q\left(f^*\right)-q_0\right)^2}{2L^2}\right)\right)\\
    &\leq \left|\Gamma_1\right|\exp\left(-n\left(c^D+\frac{\left(\max\left\{c^S,q_0\right\}-q_0\right)^2}{2L^2}\right)\right)\\
    &\leq \left(n+1\right)^m\exp\left(-n\left(c^D+\frac{\left(\left(c^S-q_0\right)^+\right)^2}{2L^2}\right)\right).
\end{align*}

\subsection{Proof of Theorem \ref{theorem: misdetection}}
Given \(c\) such that \(\left(c^S-q_0\right)^++\sqrt{2L^2c^D}<\underline{q}-q_0\), \(\Gamma_0= \left\{f\in\mathcal{P}^n\middle|q\left(f\right)<c^S\ \text{or}\ I\left(f\middle\|f^*\right)<c^D\right\}\) and for all \(f\in\Gamma_0\), either \(q(f)<c^S\) or \(q(f)\geq c^S\), \(I\left(f\middle\|f^*\right)<c^D\) and
\begin{align*}
    q(f)&\leq q\left(f^*\right)+\left|q(f)-q\left(f^*\right)\right|\\
    &\leq \max\left\{c^S,q_0\right\}+L\sqrt{2I\left(f\middle\|f^*\right)}\\
    &\leq \max\left\{c^S,q_0\right\}+\sqrt{2L^2c^D}.
\end{align*}
Thus, for any \(f_1\),
\begin{align*}
    P_{f_1,1}\left(\hat{f}_{X_1^n}\in\Gamma_0\right)&= \sum_{f\in\Gamma_0}P_{f_1,1}\left(\hat{f}_{X_1^n}=f\right)\\
    &\leq \sum_{f\in\Gamma_0}\exp\left(-nI\left(f\middle\|f_1\right)\right)\\
    &\leq \sum_{f\in\Gamma_0}\exp\left(-n\frac{\left(q(f)-q(f_1)\right)^2}{2L^2}\right)\\
    &\leq \sum_{f\in\Gamma_0}\exp\left(-n\frac{\left(q_1-\max\left\{c^S,q_0\right\}-\sqrt{2L^2c^D}\right)^2}{2L^2}\right)\\
    &\leq \left|\Gamma_0\right|\exp\left(-n\left(\frac{q_1-\max\left\{c^S,q_0\right\}}{\sqrt{2}L}-\sqrt{c^D}\right)^2\right)\\
    &\leq \left(n+1\right)^m\exp\left(-n\left(\frac{q_1-q_0-\left(c^S-q_0\right)^+}{\sqrt{2}L}-\sqrt{c^D}\right)^2\right)
\end{align*}
and
\begin{align*}
    &\sup_{f_1}P_{f_1,1}\left(\hat{f}_{X_1^n}\in\Gamma_0\right)\\
    &\leq \left(n+1\right)^m\exp\left(-n\left(\frac{\underline{q}-q_0-\left(c^S-q_0\right)^+}{\sqrt{2}L}-\sqrt{c^D}\right)^2\right).
\end{align*}

\subsection{Proof of Corollary \ref{corollary: tcd}}
Consider the probability of false alarm. The alarm time can only occur at \(k'(l)=\frac{n+1}{2}l\) for positive integers \(l=1,\dots\) and each window has \(\frac{n+1}{2}\) samples i.i.d. with \(f_0\). Then,
\begin{align*}
    1-P_\infty\left(k\leq t_I<k+d_\alpha\right)&\geq \prod_{l=\left\lceil\frac{2k}{n+1}\right\rceil}^{\left\lfloor\frac{2(k+n_\alpha))}{d+1}\right\rfloor}1-P_\infty\left(t_I=k'(l)\right)\\
    &= \left(1-P_\infty\left(t_I=k'(1)\right)\right)^{\left\lceil\frac{2n_\alpha}{n+1}\right\rceil}\\
    &\geq (1-\exp(-\nu n))^{\left\lceil\frac{2n_\alpha}{n+1}\right\rceil}
\end{align*}
where we have used Theorem \ref{theorem: false-alarm}. Thus,
\begin{align*}
    \sup_kP_\infty\left(k\leq t_I<k+n_\alpha\right)&\leq 1-(1-\exp(-\nu n))^{\left\lceil\frac{2n_\alpha}{n+1}\right\rceil}
\end{align*}
Consider the probability of misdetection. Since the window size is \(\frac{n+1}{2}\) the IPT with fixed size and rolling window will have at least one window where samples are i.i.d. with \(f_1\). Then, the upper bound is a direct consequence of Theorem \ref{theorem: misdetection}.

\subsection{Proof of Theorem \ref{theorem: arl}}
Let us express \(E_\infty t_I\) conditional on \(t_S\).
\begin{align*}
    E_\infty t_I&= E_\infty\left(E_\infty\left(t_I\middle|t_S\right)\right)\\
    &= E_\infty\left(P_\infty\left(t_I=t_S\middle|t_S\right)t_S+P_\infty\left(t_I\neq t_S\middle|t_S\right)\left(t_S+E_\infty\left(t_I\right)\right)\right)\\
    &= E_\infty\left(P_\infty\left(t_I=t_S\middle|t_S\right)t_S+P_\infty\left(t_I\neq t_S\middle|t_S\right)\left(t_S+E_\infty t_I\right)\right)\\
    &= E_\infty\left(t_S+P_\infty\left(t_I\neq t_S\middle|t_S\right)E_\infty t_I\right)\\
    &= E_\infty t_S+E_\infty P_\infty\left(t_I\neq t_S\middle|t_S\right)E_\infty t_I\\
    &= \frac{E_\infty t_S}{E_\infty P_\infty\left(t_I=t_S\middle|t_S\right)}\\
    &= \frac{E_\infty t_S}{E_\infty P_\infty\left(D_{t_S}\geq c_n^D\middle|t_S\right)}
\end{align*}
Assume \(c_n^D=c^D\) for \(n>(1+\rho)\frac{c^S}{\underline{q}}\), then, we can upper bound the denominator as follows.
\begin{align*}
    P_\infty\left(D_{t_S}\geq c_{n_{t_S}}^D\middle|t_S\right)&= \sum_{n=1}^{t_S}P_\infty\left(n_{t_S}=n\middle|t_S\right)P_\infty\left(D_{t_S}\geq c_n^D\middle|n_{t_S}=n,t_S\right)\\
    &= P_\infty\left(n_{t_S}\leq(1+\rho)\frac{c^S}{\underline{q}}\middle|t_S\right)+\sum_{n>(1+\rho)\frac{c^S}{\underline{q}}}^{t_S}P_\infty\left(n_{t_S}=n\middle|t_S\right)P_\infty\left(D_{t_S}\geq c^D\middle|n_{t_S}=n,t_S\right)\\
    &\leq P_\infty\left(n_{t_S}\leq(1+\rho)\frac{c^S}{\underline{q}}\middle|t_S\right)+\sum_{n>(1+\rho)\frac{c^S}{\underline{q}}}^{t_S}P_\infty\left(I\left(\hat{f}_{X_{t_S-n+1}^{t_S}}\middle\|f_n^*\right)\geq c^D\middle|n_{t_S}=n,t_S\right)\\
    &\leq \max_{n\leq(1+\rho)\frac{c^S}{\underline{q}}}P_\infty\left(q\left(\hat{f}_{X_{t_S-n+1}^{t_S}}\right)\geq\frac{c^S}{n}\middle|t_S\right)+\sum_{n>(1+\rho)\frac{c^S}{\underline{q}}}^{t_S}(n+1)^m\exp\left(-nc^D\right)\\
    &\leq \max_{n\leq(1+\rho)\frac{c^S}{\underline{q}}}P_\infty\left(q\left(\hat{f}_{X_{t_S-n+1}^{t_S}}\right)-q_0\geq\frac{c^S}{n}-q_0\middle|t_S\right)+\sum_{n>(1+\rho)\frac{c^S}{\underline{q}}}^{t_S}(n+1)^m\exp\left(-nc^D\right)\\
    &\leq \max_{n\leq(1+\rho)\frac{c^S}{\underline{q}}}P_\infty\left(\norm{\hat{f}_n-f_0}_1\geq\frac{\frac{c^S}{n}-q_0}{L}\middle|t_S\right)+\sum_{n>(1+\rho)\frac{c^S}{\underline{q}}}^{t_S}(n+1)^m\exp\left(-nc^D\right)\\
    &\leq \max_{n\leq(1+\rho)\frac{c^S}{\underline{q}}}2^m\exp\left(-\frac{n}{2}\left(\frac{\frac{c^S}{n}-q_0}{L}\right)^2\right)+\sum_{n>(1+\rho)\frac{c^S}{\underline{q}}}^{t_S}(n+1)^m\exp\left(-nc^D\right)\\
    &\leq 2^m\exp\left(-\frac{2\left|q_0\right|}{L^2}c^S\right)+\sum_{n>(1+\rho)\frac{c^S}{\underline{q}}}^{t_S}\left((1+\rho)\frac{c^S}{\underline{q}}+1\right)^{\frac{m\underline{q}}{(1+\rho)c^S}n}\exp\left(-nc^D\right)
\end{align*}
since \(n+1\leq\left((1+\rho)\frac{c^S}{\underline{q}}+1\right)^{\frac{\underline{q}}{(1+\rho)c^S}n}\) for \(n>(1+\rho)\frac{c^S}{\underline{q}}\). Also, w.l.o.g., we assumed \(\left|q_0\right|\geq(1+\rho)\underline{q}\). Then,
\begin{align*}
    E_\infty P_\infty\left(D_{t_S}\geq c_{n_{t_S}}^D\middle|t_S\right)&\leq 2^m\exp\left(-\frac{2\left|q_0\right|}{L^2}c^S\right)+\sum_{n>(1+\rho)\frac{c^S}{\underline{q}}}\left((1+\rho)\frac{c^S}{\underline{q}}+1\right)^{\frac{m\underline{q}}{(1+\rho)c^S}n}\exp\left(-nc^D\right)\\
    &\leq 2^m\exp\left(-\frac{2\left|q_0\right|}{L^2}c^S\right)+\frac{\left((1+\rho)\frac{c^S}{\underline{q}}+1\right)^m\exp\left(-(1+\rho)\frac{c^D}{\underline{q}}c^S\right)}{1-\left((1+\rho)\frac{c^S}{\underline{q}}+1\right)^{\frac{m\underline{q}}{(1+\rho)c^S}}\exp\left(-c^D\right)}\\
    &\to \exp\left(-\frac{2\left|q_0\right|}{L^2}c^S\right)
\end{align*}
for \(c^D\geq \frac{2\left|q_0\right|\underline{q}}{(1+\rho)L^2}\). Next, we bound lower bound \(E_\infty t_S\). We assume \(q\) is the mean operator and \(S_k\) is the random walk that is bounded below by zero. Denote the number of zero crossings as \(N_0\) and the latest upper or lower threshold crossing time as \(k^*\), then,
\begin{align*}
    E_\infty t_S&= E_\infty E_\infty\left(t_S\middle|N_0\right)\\
    &= E_\infty\left(N_0E_\infty\left(k^*\middle|S_{k^*}\leq 0\right)+E_\infty\left(k^*\middle|S_{k^*}\geq\lambda_1\right)\right)\\
    &= E_\infty N_0E_\infty(k^*|S_{k^*}\leq 0)+E_\infty(k^*|S_{k^*}\geq\lambda_1)
\end{align*}
The random walk has i.i.d. and bounded steps. Therefore its steps has moments of all order and we can apply Wald's identity. For every \(v\in\mathbb{R}\),
\begin{align*}
    E_\infty\left(\exp(vS_{k^*}-k^*\log\psi(v))\right)&= 1
\end{align*}
and
\begin{align*}
    P_\infty(S_{k^*}\geq\lambda_1)&\leq \exp(-v^*c^S)
\end{align*}
Thus,
\begin{align*}
    E_\infty t_S&= \left(\exp(v^*c^S)-1\right)E_\infty(k^*|S_{k^*}\leq 0)+E_\infty(k^*|S_{k^*}\geq c^S)\\
    &\geq \exp(v^*c^S)
\end{align*}
Finally we obtain,
\begin{align*}
    E_\infty t_I&\geq \frac{\exp(v^*c^S)}{2^m\exp\left(-\frac{2\left|q_0\right|}{L^2}c^S\right)+\frac{\left((1+\rho)\frac{c^S}{\underline{q}}+1\right)^m\exp\left(-(1+\rho)\frac{c^D}{\underline{q}}c^S\right)}{1-\left((1+\rho)\frac{c^S}{\underline{q}}+1\right)^{\frac{m\underline{q}}{(1+\rho)c^S}}\exp\left(-c^D\right)}}\\
    &\to \exp\left(\left(v^*+\frac{2\left|q_0\right|}{L^2}\right)c^S\right)
\end{align*}

\subsection{Proof of Lemma \ref{lemma: stopping}}
Let \(j=0,\dots\) be arbitrary. W.l.o.g. assume \(t_I\geq t_S^{(j)}\). If \(D_{t_S^{(j)}}\geq c^D\), we have \(t_I=t_S^{(j)}\). Else, the stopping time is restarted at \(t_S^{(j)}+1\) and \(t_I=t_S^{(j)}+t_I^{t_S^{(j)}+1}\). In any case, \eqref{equation: stopping} is satisfied.

Let \(j=1,\dots,\) and \(t_S^{(j-1)}<k\leq t_S^{(j)}\) be arbitrary. If \(\tau_{k+t_S^k-1}>k\), then there exists \(k'\) such that \(k<k'\leq k+t_S^k-1\) and \(\tau_{k+t_S^k-1}=\tau_{k'}=k'\). Thus, \(t_S^{(j)}\leq k'\leq k+t_S^k-1\). Else, \(\tau_{k+t_S^k-1}\leq k\)
\begin{align*}
    S_{k+t_S^k-1}&= \max_{\tau_{k+t_S^k-1}\leq i\leq k+t_S^k}Q(i,k)\\
    &\leq \max_{k\leq i\leq k+t_S^k}Q(i,k)\\
    &= S_{t_S^k}^k\geq c^S.
\end{align*}
Thus, \(t_S^{(j)}\leq k+t_S^k-1\).

Finally, let \(j^*=\min\left\{j\middle|t_S^{(j)}\geq t_1\right\}\). Then, using \eqref{equation: stopping} and \eqref{equation: auxilary stopping} with \(j=j^*\) and \(k=t_1\) we get
\begin{align*}
    t_I&\leq t_S^{(j^*)}+t_I^{t_S^{(j^*)}+1}\\
    &\leq t_1+t_S^{t_1}-1+t_I^{t_S^{(j^*)}+1}.
\end{align*}

\subsection{Proof of Lemma \ref{lemma: wadd}}
For any \(f_1, t_1\) and \(X_1^{t_1-1}\),
\begin{align*}
    \left(t_I-t_1+1\right)^+&\leq \left(t_1+t_S^{t_1}-1+t_I^{t_S^{j^*}+1}-t_1+1\right)^+\\
    &= t_S^{t_1}+t_I^{t_S^{(j^*)}+1}
\end{align*}
Therefore,
\begin{align*}
    E_{f_1,t_1}\left(\left(t_I-t_1+1\right)^+\middle|X_1^{t_1-1}\right)&\leq E_{f_1,t_1}\left(t_S^{t_1}+t_I^{t_S^{(j^*)}+1}\middle|X_1^{t_1-1}\right)\\
    &= E_{f_1,t_1}\left(t_S^{t_1}+t_I^{t_S^{(j^*)}+1}\right)\\
    &= E_{f_1,1}\left(t_S+t_I\right)
\end{align*}
since \(t_S^{(j^*)}\geq t_1\).
\subsection{Proof of Lemma \ref{lemma: add}}
First, we express the expected delay conditional to \(t_S\).
\begin{align*}
    E_{f_1,1}t_I&= E_{f_1,1}E_{f_1,1}\left(t_I\middle|t_S\right)\\
    &= E_{f_1,1}\left(t_S+P_{f_1,1}\left(t_I\neq t_S\right)E_{f_1,1}t_I\middle|t_S\right)\\
    &= E_{f_1,1}\left(t_S\right)+E_{f_1,1}P_{f_1,1}\left(t_I\neq t_S\middle|t_S\right)E_{f_1,1}\left(t_I\right)\\
    &= \frac{E_{f_1,1}\left(t_S\right)}{E_{f_1,1}P_{f_1,1}\left(t_I= t_S\middle|t_S\right)}
\end{align*}
The nominator is bounded using the probability that \(S_k\) does not cross the threshold until time \(t\).
\begin{align*}
    P_{f_1,1}\left(t_S>t\right)&= P_{f_1,1}\left(\inf\left\{k\middle|S_k\geq c^S\right\}>t\right)\\
    &= P_{f_1,1}\left(\max_{1\leq k\leq t}S_k<c^S\right)\\
    &\leq P_{f_1,1}\left(Q(1,t)\right)\\
    &\leq P_{f_1,1}\left(tq\left(\hat{f}_{X_1^t}\right)<c^S\right)\\
    &\leq P_{f_1,1}\left(q\left(\hat{f}_{X_1^t}\right)<\frac{c^S}{t}\right)\\
    &\leq P_{f_1,1}\left(q\left(\hat{f}_{X_1^t}\right)-q_1<\frac{c^S}{t}-q_1\right)\\
    &\leq P_{f_1,1}\left(\left|q\left(\hat{f}_{X_1^t}\right)-q_1\right|>q_1-\frac{c^S}{t}\right)\\
    &\leq P_{f_1,1}\left(L\norm{\hat{f}_{X_1^t}-f_1}_1>q_1-\frac{c^S}{t}\right)\\
    &= P_{f_1,1}\left(\norm{\hat{f}_{X_1^t}-f_1}_1>\frac{q_1-\frac{c^S}{t}}{L}\right)\\
    &\leq 2^m\exp\left(-\frac{t}{2}\frac{q_1-\frac{c^S}{t}}{L}\right)\\
    &\leq 2^m\exp\left(-\frac{q_1t-c^S}{2L}\right)
\end{align*}
Then, for any \(\delta>0\),
\begin{align*}
    E_{f_1,1}t_S&= \sum_{t=0}^\infty P_{f_1,1}\left(t_S>t\right)\\
    &= (1+\delta)\frac{c^S}{q_1}+\sum_{t>(1+\delta)\frac{c^S}{q_1}} P_{f_1,1}\left(t_S>t\right)\\
    &\leq (1+\delta)\frac{c^S}{q_1}+\sum_{t>(1+\delta)\frac{c^S}{q_1}} 2^m\exp\left(-\frac{q_1t-c^S}{2L}\right)\\
    &= (1+\delta)\frac{c^S}{q_1}+2^m\frac{\exp\left(-\frac{\delta c^S}{2L}\right)}{1-\exp\left(-\frac{q_1}{2L}\right)}
\end{align*}
The denominator is bounded using the fact that \(t_S\leq (1+\rho)\frac{c^S}{q_1}\) with high probability.
\begin{align*}
    P_{f_1,1}\left(t_I=t_S\middle|t_S\right)&\geq P_{f_1,1}\left(n_{t_S}\leq(1+\rho)\frac{c^S}{\underline{q}}\middle|t_S\right)P_{f_1,1}\left(t_I=t_S\middle|n_{t_S}\leq(1+\rho)\frac{c^S}{\underline{q}},t_S\right)\\
    &= P_{f_1,1}\left(n_{t_S}\leq(1+\rho)\frac{c^S}{\underline{q}}\middle|t_S\right)\\
    &\geq \mathbb{1}\left(t_S\leq(1+\rho)\frac{c^S}{\underline{q}}\right)
\end{align*}
since \(c^D_n=0\) and \(t_I=t_S\) for \(n_{t_S}\leq(1+\rho)\frac{c^S}{\underline{q}}\). So,
\begin{align*}
    E_{f_1,1}P_{f_1,1}\left(t_I= t_S\middle|t_S\right)&\geq E_{f_1,1}\mathbb{1}\left(t_S\leq(1+\rho)\frac{c^S}{\underline{q}}\right)\\
    &= P_{f_1,1}\left(t_S\leq(1+\rho)\frac{c^S}{\underline{q}}\right)\\
    &= P_{f_1,1}\left(\max_{k\leq(1+\rho)\frac{c^S}{\underline{q}}}S_k\geq c^S\right)\\
    &\geq P_{f_1,1}\left(S_{(1+\rho)\frac{c^S}{\underline{q}}}\geq c^S\right)\\
    &\geq P_{f_1,1}\left(Q\left(1,(1+\rho)\frac{c^S}{\underline{q}}\right)\geq c^S\right)\\
    &= P_{f_1,1}\left((1+\rho)\frac{c^S}{\underline{q}}q\left(\hat{f}_{X_1^{(1+\rho)\frac{c^S}{\underline{q}}}}\right)\geq c^S\right)\\
    &= P_{f_1,1}\left(q\left(\hat{f}_{X_1^{(1+\rho)\frac{c^S}{\underline{q}}}}\right)\geq\frac{\underline{q}}{1+\rho}\right)\\
    &= 1-P_{f_1,1}\left(q\left(\hat{f}\right)<\frac{\underline{q}}{1+\rho}\right)\\
    &= 1-P_{f_1,1}\left(\left|q\left(\hat{f}\right)-q\left(f_1\right)\right|>q_1-\frac{\underline{q}}{1+\rho}\right)\\
    &\geq 1-P_{f_1,1}\left(\norm{\hat{f}-f_1}_1>\frac{q_1-\frac{\underline{q}}{1+\rho}}{L}\right)\\
    &\geq 1-2^m\exp\left(-(1+\rho)\frac{c^S}{\underline{q}}\frac{\left(\frac{q_1-\frac{\underline{q}}{1+\rho}}{L}\right)^2}{2}\right)\\
    &\geq 1-2^m\exp\left(-\frac{\rho^2\underline{q}}{2(1+\rho)L^2}c^S\right)
\end{align*}
As \(c^S\to\infty\),
\begin{align*}
    E_{f_1,1}t_I&= \frac{E_{f_1,1}\left(t_S\right)}{E_{f_1,1}P_{f_1,1}\left(t_I= t_S\middle|t_S\right)}\\
    &\leq \frac{(1+\delta)\frac{c^S}{q_1}+2^m\frac{\exp\left(-\frac{-\delta c^S}{2L}\right)}{1-\exp\left(-\frac{q_1}{2L}\right)}}{1-2^m\exp\left(-\frac{\rho^2\underline{q}}{2(1+\rho)L^2}c^S\right)}\to\frac{c^S}{q_1}
\end{align*}
since \(\delta>0\) is arbitrary.

\subsection{Proof of Theorem \ref{theorem: wadd}}
Using Lemma \ref{lemma: wadd} and \ref{lemma: add}, as \(c^S\to\infty\),
\begin{align*}
    WADD(t_1)&= \sup_{f_1,t_1}\esssup_{X_1^{t_1-1}} E_{f_1,t_1}\left(\left(t_I-t_1+1\right)^+\middle|X_1^{t_1-1}\right)\\
    &\leq \sup_{f_1,t_1}\esssup_{X_1^{t_1-1}}E_{f_1,1}\left(t_S+t_I\right)\\
    &\leq \sup_{f_1}E_{f_1,1}\left(t_S+t_I\right)\\
    &\leq \sup_{f_1}\frac{2c^S}{q_1}\\
    &\leq \frac{2c^S}{\underline{q}}
\end{align*}

\bibliographystyle{IEEEtran}
\bibliography{bibliography.bib}

\end{document}